# Five Ps: Leverage Zones towards Responsible AI

**Ehsan Nabavi\*, Chris Browne**

*Responsible Innovation Lab, Australian National Centre for Public Awareness of Science, The Australian National University, Canberra, Australia*

April 7, 2022

## ABSTRACT

There is a growing debate amongst academics and practitioners on whether interventions made, thus far, towards Responsible AI would have been enough to engage with root causes of AI problems. Failure to effect meaningful changes in this system could see these initiatives to not reach their potential and lead to the concept becoming another buzzword for companies to use in their marketing campaigns.

We propose that there is an opportunity to improve the extent to which interventions are understood to be effective in their contribution to the change required for Responsible AI. Using the notions of 'leverage zones' adapted from the Systems Thinking literature, we suggest a novel approach to evaluate the effectiveness of interventions, to focus on those that may bring about the real change that is needed.

In this paper we argue that insights from using this perspective demonstrate that the majority of current initiatives taken by various actors in the field, focus on low-order interventions, such as short-term fixes, tweaking algorithms and updating parameters, absent from higher-order interventions, such as redefining the system's foundational structures that govern those parameters, or challenging the underlying purpose upon which those structures are built and developed in the first place (high-leverage).

This paper presents a conceptual framework called the Five Ps to identify interventions towards Responsible AI and provides a scaffold for transdisciplinary question-asking to improve outcomes towards Responsible AI.

## 1. Introduction

People today are increasingly aware of how ingrained Artificial Intelligence (AI) already is in their daily lives—whether it determines what appears in a playlist or suggests potential partners to date—rather than in some distant future. While these seemingly low-risk examples can feel like magic to the user, many more technological advances are also underway that delegate more significant control over decision-making to AI-systems, such as in driving [1], educating [2], judicial applications [3], and providing health care [4].

However, the outputs from these systems can inadvertently erode the shared values of society, such as fairness, justice, safety, security, and accountability. The problems that AI is employed to solve can often exacerbate other societal problems, such as loss of privacy through increased surveillance [5], and policy decisions that increase social and economic inequality [6-8]. Recent examples of AI failures and their lack of transparency and traceability have raised disconcerting questions about the 'dark side'[9] of AI use, and the way these systems are developed and deployed [10].

Advances in digital technology, along with debates about biased algorithms and ethical and regulatory challenges of autonomous systems, underscore the fact that AI management is more of a social and political issue rather than an engineering challenge [11]. This realization has caused research and industry actors to take non-technical aspects of AI into account. This conversation has grown beyond the ethical challenges surrounding AI development and use into a broader discussion of 'Responsible AI', encompassing other topics around ethical AI, lawful AI, explainable AI (XAI), trustworthy AI, and accountable AI.

Although there is not a consensus over the meaning and implications of the notion of 'responsibility' when it is applied to AI-systems[12], there is a growing interest to explore the development and use of AI systems from the lens of Responsible AI. Applications range from in fields such as health [13,14], finance [15], urban studies [16], conservation science [17], marketing [18], and military affairs [19], to more specific cases such as COVID-19 [20].

The notion of Responsible AI is not limited to research. As of January 2021, OECD AI Policy Observatory tracks more than 300 AI policy initiatives around the globe in the Responsible AI landscape [21]. The major AI companies have launched their self-regulatory Responsible AI programs, through building tools and software to translate high-level principles such as fairness, explainability, and accountability and use them across engineering groups and clients [22]. Standardization bodies such as ISO, IEEE and NIST also offer guidance by publishing standards and frameworks to support the responsible development of AI.

Although transdisciplinary approaches can help us to understand and navigate this socio-technical challenge, the dominant discourses address AI problems are from disciplinary perspectives, predominantly from computer science and engineering. Even within Responsible AI, researchers and practitioners tend to approach the topic from a narrowly disciplinary perspective and develop solutions based on their own epistemological strategies. For AI systems that are perceived as irresponsible, the focus is often addressing visible gaps and tangible problems with quick fixes and technical improvement (particularly in areas such as robustness, privacy, and fairness where technical fixes seem feasible)[23,24], rather than examining the drivers of design, development, and deployment. This is the gap that needs to be addressed if we are to ensure transformation of responsible AI systems.

Further, companies seeking to improve responsibility in AI systems are limited in their capacity to realize meaningful and necessary technical, social, and environmental change. Previous attempts under the banner of Responsible AI can be described as 'ethical washing' or 'ethics theatre' [25-27], intended to (1) show their customers they are doing their best to behave ethically; and more importantly, to (2) minimize regulation. Although it can be argued that efforts thus far are 'good first steps' towards Responsible AI [22], these efforts can distract from taking a broader view of the problems inherent to Responsible AI.

For the purposes of this paper, we take a holistic, systems view of Responsible AI [28], encompassing broadly the notions of responsible, transparent, and trustworthy AI described above. We argue that although that these initiatives seek to ensure more responsible building and application of AI, they often fail to engage with how to encourage developers to approach the root causes and unintended consequences, who often rush to tweak and update existing systems with new software libraries [29]. Focusing on the engineering solution, they do little to encourage AI developers and users to question underlying assumptions about the vision and the purpose of an AI system.

For the practitioner and policymaker alike, the current body of literature on Responsible AI lacks adequate definitions and characterization of how different interactions with AI systems will lead to achieving Responsible AI. Meadows [30] proposes the framework of leverage points, which summarizes the relative power of various policies to enact change—from incremental to transformational—in a system.

In the following sections we adapt Meadows' framework to the current debates in AI to help realize the transformational change needed towards Responsible AI. We conclude with a short discussion on the advantages of our proposed approach to advance theoretical and practical discussions on responsible approaches for AI.

## 2. Leverage zones to realize change

Meadows identifies twelve leverage points that has been adapted into research and practical work in various disciplines concerning complex socio-technical systems, from food and energy system [31], to climate change [32], and health [33]. These leverage points represent at an abstract level some common places to intervene within a system to effect change.

Here we adapt the leverage point framework categorized around two domains and four zones. Figure 1 shows a graphical depiction of the Five Ps framework. The two domains—*Problem and Response*—are represented by a triangle divided into two with the Problem Domain on the left and Response Domain on the right. The horizontal axis represents the relative magnitude of 'effort' and reward for intervening in each of the four zones, shown on the vertical axis in increasing order of 'leverage', from top to bottom: *Parameter, Process, Pathway, Purpose*. We describe this framework as the 'Five Ps', which includes situating the Problem at the right level, and then considering places to intervene in the system in each of the four zones.

The Five Ps is a method for considering and analyzing what the response of a given intervention might be. In this case, we are considering how different initiatives in Responsible AI are conceived in the Problem Domain, and then enacted in the four zones (i.e. Parameter, Process, Pathway, Purpose). This framework recognizes that a problem can be attributed to various parts of a system and, depending on how the problem is framed, different interventions will be chosen, each leading to a different response. To illustrate, the misclassification problem that arises in an AI model could be seen as a Parameter problem that can be resolved by improving parameters within the algorithm. However, the problem could also be seen as a Purpose problem, which could bring into question the paradigm and assumptions from which the algorithm was created in the first place.

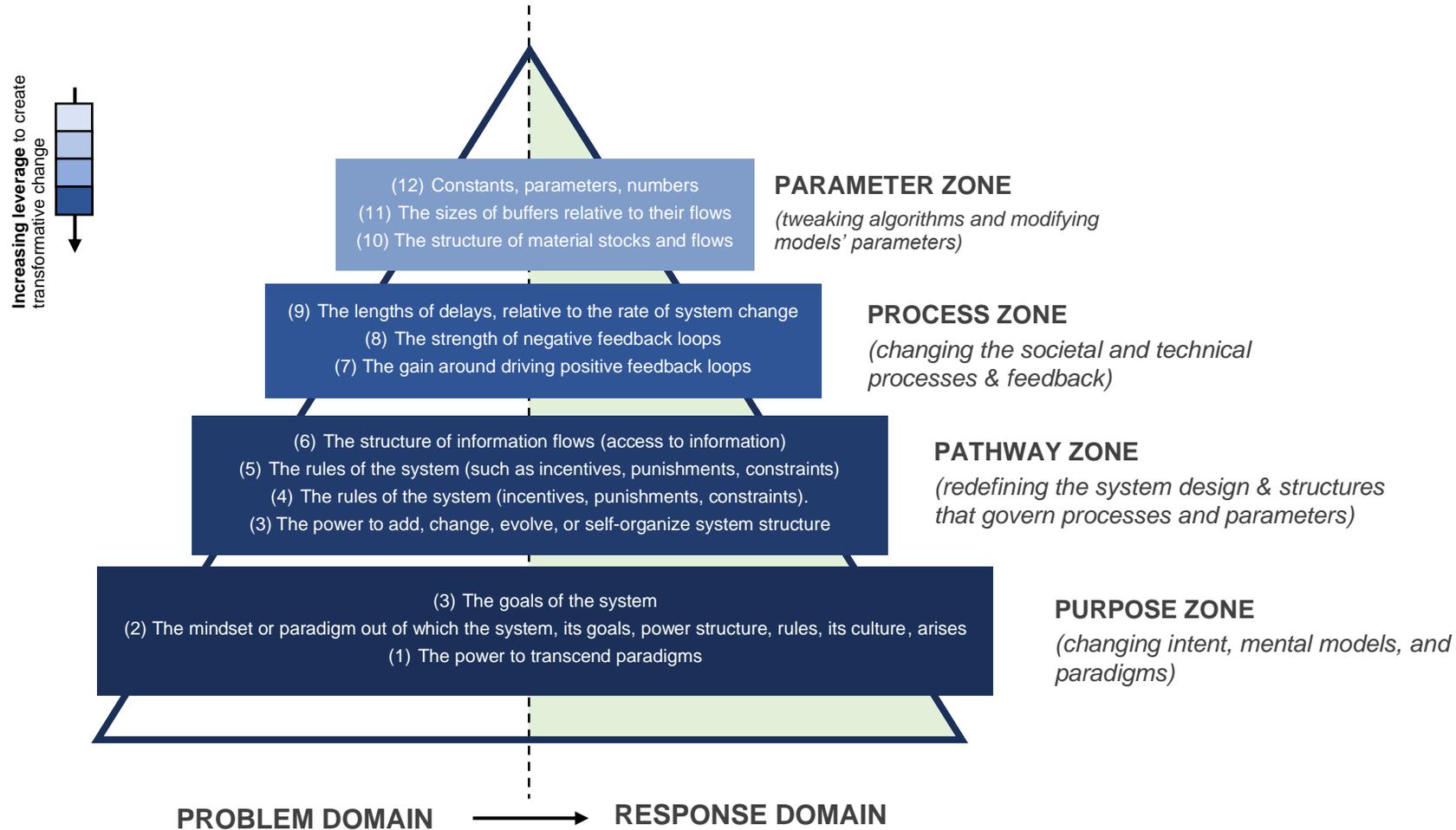

**Figure 1** *The diagram represents the 'leverage zones' within which interventions are made. The twelve leverage points on the pyramid are the places where Meadow (1999) suggests for interventions in a given system.*

To illustrate the domains and zones within the Five Ps, we will describe each briefly in relation to Responsible AI.

Problems identified in the *Parameter* zone are tractable (modifiable, mechanistic) characteristics of an AI system that are commonly targeted by AI developers to improve the responsibility of AI. They are typically smaller visible flaws that are usually addressed through engineering solutions such as tweaking algorithms and parameters. The effort to fix these is small, and changes in this zone are incremental and may have a negligible effect on the problem's underlying structure or dynamics. They are important markers of the problem, but they are often symptomatic and not the root cause of the problem.

Problems identified in the *Process* zone consider the wide range of interactions between the feedback elements of an AI system that drive the internal dynamics, including social and technical processes associated with how the AI is designed, built, and deployed. This might include activities that speed up development times, or actively responding to emerging trends in the data. Changes in this zone are likely to result in resolving issues as they emerge or amplifying the effect of assumptions.

Problems identified in the *Pathway zone* consider the ways through which information flows, the rules are set, and the power is organized. For example, improving transparency of how algorithms are employed, the governance or legislation of their use, or putting the ownership of data back into the consumer's hands. These changes are structural to the system that allows the AI to operate, and result in establishing new patterns of behavior and agency.

Issues identified in the *Purpose* zone have the most potential to affect change in a system. These relate to the norms, values, goals, and worldviews of AI developers that are embodied in the system. It includes the underpinning paradigms based on which the system is imagined, and the ability to transform entirely and imagine new paradigms. Framing perceived problems in this zone serves to act as a compass to guide the developers to align with the fundamental purpose of the system.

The Five Ps—*problem, parameter, process, pathway, and purpose*— characterize five ways we can begin to conceptualize our journey towards Responsible AI. Zones within the Five Ps are interrelated, and scale and reach also plays a role in the extent to which the system's behavior changes. These Five Ps are not part of a fixed hierarchy of change but serve as a conceptual tool to categorize strategies to effect change in a system. In the following sections, we consider the Five Ps as an analytical tool and how it could be used as a planning tool to assess current interventions in Responsible AI.

## 3.  The Five Ps as an analytical tool

Reviewing the ongoing attempts to address Responsible AI, it is common to see that activities are commonly conceptualized, defined, and implemented in the Parameters and Process zones. Many initiatives frame the challenge of Responsible AI as the problem of technical and design flaws requiring engineering fixes or a better design process [29,34]. The rational is that complex concepts and high-level principles need to be simplified so as to be tangible and computable. The result had been studies that examined isolated factors related to the principles, such as improving model explainability[35] and reducing biases[36] . Change at these levels typically result in incremental change and allow business as usual.

There are opportunities to radically shift a movement to Responsible AI by effecting changes in Pathway zone, such as high-level design and structures, and the challenging questions about the underlying assumptions, visions, and the foundational purpose of the system as in Purpose zone. This particularly happens when Responsible AI is understood as the microcosm of cultural and political challenges faced in society[25], beyond technical and design issues.

To illustrate, consider an AI system that is used in a social media company causing misinformation and extremism—similar to the one Facebook is currently experiencing. In a move towards Responsible AI, the company views the problem at the Parameters zone, and creates tools and tweaks algorithms to analyze and address the biases in order to fix the models that come out of them. Another response from the company, could be creating software tools to translate principles of responsible AI, such as fairness, explainability, and accountability to improve the models [22].By taking these measures, the company seeks to control misinformation in its content-moderation models across the platform, which potentially leads to an improved user experience.

These interventions could be described as 'technological solutionism', built on a premise that the challenge of responsibility is a challenge of fixing a design flaw in the algorithms [25,37-39]. In this view, the efforts for quantifying, computing, or ma thematizing responsibility are perceived as an apparatus for creating a technocratic rather than democratic solutions, and fixes tend to be short-term and could be described as 'tweaks reaction'.

In our example, although visible content moderation could improve, the paradigm under which the platform operates remains unchanged. If the company's business model only concerns with maximizing engagement, tweaking algorithms will then have no direct impact on misinformation circulation— because the AI models that governs the interactions will continue to reward inflammatory content (e.g., controversy, misinformation, and extremism), and operate on structures that systematically reduce the diversity of viewpoints that users are exposed to.

Further, engaging changes that undermine the company's paradigms are unlikely to be supported. For example, a for-profit company is unlikely to support initiatives that have potential to reduce revenue streams [40,41].

The same company could consider the problem in the Process zone, by intentionally promoting diversity and inclusion in development teams, publishing new professional guidelines and promoting training opportunities. As more diverse views are involved in the development of the model, assumptions are questioned and resolved during the development cycle. This would likely see first-order change, adjusting and adapting practices to changes in the operating environment (see Figure 2).

Extending this, the company could initiate reform in the Pathway zone to achieve second-order change through 'restructures' and 'redesigns'. For example, this could include initiating governance structures within their firm for Responsible AI, such as ethical review boards or introducing new roles and responsibilities for assuring AI products and processes are ethical and aligned with AI principles the company abides by. Collective partnerships can also focus discussion on the development of design principles, guidelines, and best practices for AI [42]. However, a unified and strong regulation does not yet exist which can establish fiduciary duties to the public, and that implies the societies can just hope that reputational risks or company's own values and standards may create more responsible approaches towards AI development and use[43]. Further, partnerships thus far have produced "vague, high-level principles and value statements which promise to be action-guiding, but in practice provide few specific recommendations and fail to address fundamental normative and political tensions embedded in key concepts for example, in fairness and privacy" [25].

Finally, in the Purpose zone, the same company could deploy resources to move to third-order, transformative change by 'reconsidering' or 'redefining' the purpose of their system. In the example of social media company, this could be a change in purpose from maximizing engagement to activities such as truth-seeking or social cohesion. There are, for example, several experimental products, such as a platform called Polis, that highlight diverse views and work towards maximizing 'consensus' rather than engagement, and thereby fundamentally changing the goal of the system.

This demonstrates that there are often multiple interactions between leverage zones which can be studied for evaluating the intervention's effectiveness. These zones are not discrete, and for effective implementation of change, we should consider the interactions required across an entire system for change in the deeper leverage zones.

Despite efforts to move towards Responsible AI, many of these initiatives, particularly those conducted at the corporate level, have been characterized by critics as 'ethics washing', where industry adopts 'appearances of ethical behavior' for self-serving purposes (for example, to reduce regulatory requirements or maintain self-regulation) [23,26,44,45]. On the other hand, it is argued that steps forward in the right direction, however small, are welcome, and major AI companies who have had concrete plans and actions help the industry move towards responsible AI [22].

As an analytic tool, the Five Ps can be used to view the relative strength of interventions towards Responsible AI. In the following section, we look at how the Five Ps can also be used as a planning tool by those seeking to deliver Responsible AI.

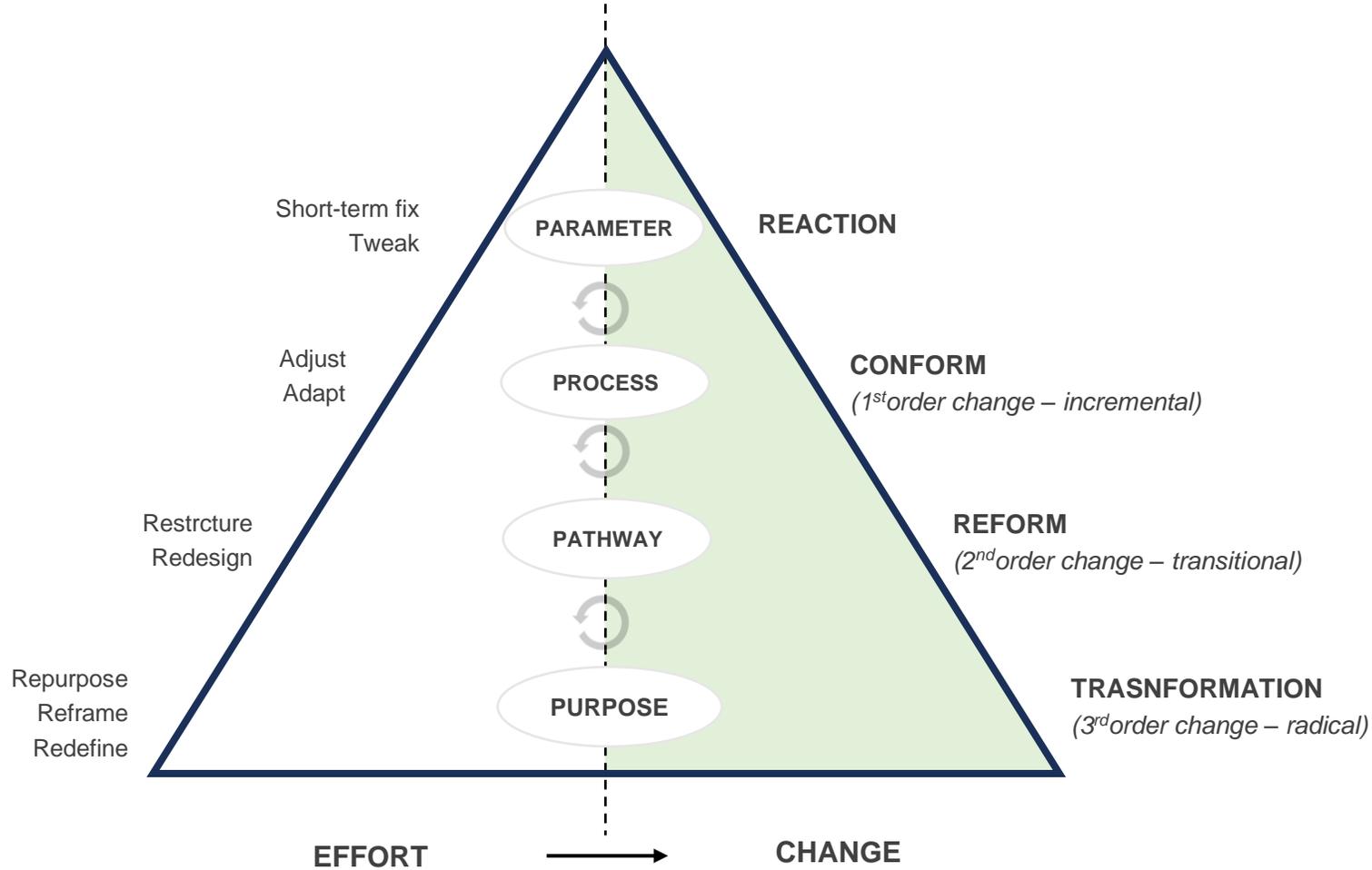

**Figure 2** *Schematic illustration of the leverage zones, showing their differences in terms of 'efforts' they are needed and the type of 'change' they bring about. Feedback loops indicate interactions that may happen between and among different leverage zones.*

## 4. The Five Ps as a planning tool

Efforts towards Responsible AI thus far have predominantly been narrow in focus, with leverage zones chosen that have failed to address challenges 'deeper' questions about the governing rules, structure, business model, and purpose. To move towards Responsible AI, we argue that interventions should be seen and studied in a holistic manner, not in isolation, to avoid missing linkages between the leverage zones, to prioritize competing efforts, to consider the narrow and broad consequences, and to plan in the short and long term.

As a planning tool, the Five Ps can be used to prompt consideration of a given problem at multiple levels to achieve the desired level of 'response'. In Table 1, we provide a set of questions for each leverage zone that could be considered when considering a potential intervention. These questions should be seen as a general set of considerations; they are not exhaustive, and should be tailored to the situation at hand. By proactively considering questions that address systems-level concerns within each of the leverage zones, the appropriate leverage zone of the problem can be properly assessed, and possible synergies and contradictions that might arise can be considered.

**Table 1.** *Lines of questioning on interventions for Responsible AI*

| Parameter questions | Process questions | Pathways questions | Purpose questions |
|---|---|---|---|
| How to keep the system stable with minimum change? | How should principles be drawn up and applied? | How can we change the structure of the system? | Why are we doing it? What are the goals? |
| How to step out of 'abstract' discussion by defining 'practical' actions? | How can we speed up things that are working? | What are the rules and who makes them (incentive, punishment, constraints)? | What are the fundamental assumptions behind our work? Do we need to change them? |
| How should principles be quantified? | How can we slow down things that are not working? | Who does and does not have access to what kinds of information? | How does our value system shape our work and the final product? |
| How risks and benefits can be managed through changing parameters and resources? | How can we reduce delays? | How can we share information more readily? | How our priorities drive the design choices we make? |
| What parameters need to be measured and modified? | | How can we involve users in problem solving? | Are our motivations transparent and for the public good? |
| What resources can be deployed? | | How do we know we are right? and, what is involved if things go wrong? | Who will benefit, who will lose? |
| What other impacts can we anticipate? | | | Are there other alternatives? What are they? |

By exploring these questions, the Five Ps approach allows decision-makers to better understand the scope of the change they are seeking and avoid engaging with the system in shallow leverage zones, such as focusing on AI Principles alone or developing tools and practices for explainable models. It recognizes and promotes the importance of 'question-asking' and how it can influence the shape of the pathway towards Responsible AI.

Second, it shows how focusing interventions within discrete leverage zones can precipitate in others, across various depths. The interdependencies between different leverage zones are important to be recognized and studied. Working from the deeper leverage zones shapes and limits the types of interventions available in shallower leverage zones see [46].

Third, it provides an aid for maintaining a holistic view over the challenges associated with Responsible AI, avoiding 'atomized' and 'siloed' conceptualizations in which social, technical, and governance aspects of AI systems are addressed separately, rather than elements that are tightly interacting togethers. The alternative is that we will remain in the existing paradigm which mostly overlooks the structures, norms, values, and goals underpinning the complex problems Responsible AI is facing at deeper levels. Nevertheless, given the scale of existing social and ethical problems that have emerged in relation to the AI use, there is a strong incentive for major AI companies to adopt new tools and frameworks in order to prevent the development technologies that have the possibility to cause harm [47].

And lastly, it provides a transdisciplinary context for a conversation about Responsible AI. Since AI developers come from varied disciplines (each with their own epistemic culture and ethical standards), to speak about Responsible AI, we need frameworks that can engage all stakeholders in meaningful discussions. This is particularly important as we can expect that experts interested in human and environmental aspects of AI-powered technologies are increasingly joining the conversation [11,48]. The Five Ps framework provides a new communication tool for a wide range of stakeholders to speak about their ideas and priorities for the future of AI and collaborate using qualitative and quantitative methods.

However, along with all these advantages, there are certain issues or challenges that must be addressed carefully when we use this approach. The concept of leverage zones and places to intervene in a system are in their infancy in the field of Responsible AI, and will benefit from added discussion and more research to inform where and how they can be used. The terminology has yet to be further developed and established, so we can develop methods to identify or validate leverage zones at different scales, such as temporal, institutional, network, and management factors, and societal reach, such as global, national, local levels, towards responsible AI.

## 5. Conclusion

Responsible AI needs to engage with the deep questions to find solutions that can address root causes that have led to negative outcomes in AI products and processes. As such we need to constantly reflect about whether the planned initiatives can realize the system shift required to create an environment conducive towards Responsible AI. To this end, we propose that the Five Ps framework is a useful tool to frame a conversation around the notion of 'leverage zone' as the industry takes actions towards Responsible AI.

The key advances that the Five Ps framework presents are in developing a shared understanding of: likely long-term effectiveness of proposed initiatives; interdependencies between initiatives required for long-lasting change; frames of question-asking when considering initiatives; removing barriers around silos of activity; considering the broader implications of initiatives, and; providing a transdisciplinary context for the conversation.

Further work is required to study long-term effects of decisions arising from the Five Ps zones as a planning tool. However, as we have noted in other domains, it is highly likely that any efforts towards understanding interventions towards Responsible AI at a more holistic, systems level will see benefits over taking a fragmented, siloed approach.

# References

1   Nunes, A., Reimer, B. & Coughlin, J. F.    (Nature Publishing Group, 2018).

2   Zawacki-Richter, O., Marín, V. I., Bond, M. & Gouverneur, F. Systematic review of research on artificial intelligence applications in higher education–where are the educators? *International Journal of Educational Technology in Higher Education* **16**, 1-27 (2019).

3   Cui, Y. *Artificial intelligence and judicial modernization*.  (Springer, 2020).

4   Schwalbe, N. & Wahl, B. Artificial intelligence and the future of global health. *The Lancet* **395**, 1579-1586 (2020).

5   Mitchell, A. & Diamond, L. *China's Surveillance State Should Scare Everyone*, <https://www.theatlantic.com/international/archive/2018/02/china-surveillance/552203/> (2018).

6   Caetano, T. & Simpson-Young, B. *Artificial intelligence can deepen social inequality. Here are 5 ways to help prevent this*, <https://theconversation.com/artificial-intelligence-can-deepen-social-inequality-here-are-5-ways-to-help-prevent-this-152226> (2021).

7   Walsh, M. *Algorithms are making economic inequality worse*, <https://hbr.org/2020/10/algorithms-are-making-economic-inequality-worse> (2020).

8   Perc, M., Ozer, M. & Hojnik, J. Social and juristic challenges of artificial intelligence. *Palgrave Communications* **5**, 61, doi:10.1057/s41599-019-0278-x (2019).

9   Mikalef, P., Conboy, K., Lundström, J. E. & Popovič, A. Thinking responsibly about responsible AI and 'the dark side' of AI. *European Journal of Information Systems*, 1-12, doi:10.1080/0960085X.2022.2026621 (2022).

10  Choi, C. Q. 7 Revealing Ways AIs Fail: Neural Networks can be Disastrously Brittle, Forgetful, and Surprisingly Bad at Math. *IEEE Spectrum* **58**, 42-47 (2021).

11  Nabavi, E. Why the huge growth in AI spells a big opportunity for transdisciplinary researchers. *Nature*, doi:10.1038/d41586-019-01251-1 (2019).

12  Constantinescu, M., Voinea, C., Uszkai, R. & Vică, C. Understanding responsibility in Responsible AI. Dianoetic virtues and the hard problem of context. *Ethics and Information Technology* **23**, 803-814, doi:10.1007/s10676-021-09616-9 (2021).

13 Gupta, S., Kamboj, S. & Bag, S. Role of Risks in the Development of Responsible Artificial Intelligence in the Digital Healthcare Domain. *Information Systems Frontiers*, doi:10.1007/s10796-021-10174-0 (2021).

14 Trocin, C., Mikalef, P., Papamitsiou, Z. & Conboy, K. Responsible AI for Digital Health: a Synthesis and a Research Agenda. *Information Systems Frontiers*, doi:10.1007/s10796-021-10146-4 (2021).

15 Maree, C., Modal, J. E. & Omlin, C. W. in *2020 IEEE Symposium Series on Computational Intelligence (SSCI)*. 16-21 (IEEE).

16 Yigitcanlar, T. *et al.* Responsible Urban Innovation with Local Government Artificial Intelligence (AI): A Conceptual Framework and Research Agenda. *Journal of Open Innovation: Technology, Market, and Complexity* **7**, 71 (2021).

17 Wearn, O. R., Freeman, R. & Jacoby, D. M. P. Responsible AI for conservation. *Nature Machine Intelligence* **1**, 72-73, doi:10.1038/s42256-019-0022-7 (2019).

18 Liu, R., Gupta, S. & Patel, P. The Application of the Principles of Responsible AI on Social Media Marketing for Digital Health. *Information Systems Frontiers*, doi:10.1007/s10796-021-10191-z (2021).

19 Stanley-Lockman, Z. & Trabucco, L. in *The Oxford Handbook of AI Governance* (eds Justin Bullock *et al.*) (2022).

20 Leslie, D. Tackling COVID-19 through responsible AI innovation: Five steps in the right direction. *Harvard Data Science Review* (2020).

21 Ibaraki, S. *Responsible AI Programs To Follow And Implement— Breakout Year 2021*, <https://www.forbes.com/sites/stephenibaraki/2020/12/26/responsible-ai-programs-to-follow-and-implement--breakout-year-2021/> (2021).

22 de Laat, P. B. Companies Committed to Responsible AI: From Principles towards Implementation and Regulation? *Philosophy & technology* **34**, 1135-1193 (2021).

23 Greene, D., Hoffmann, A. L. & Stark, L. in *Proceedings of the 52nd Hawaii international conference on system sciences* 10 (2019).

24 Hagendorff, T. The ethics of AI ethics: An evaluation of guidelines. *Minds and Machines* **30**, 99-120 (2020).

25 Mittelstadt, B. Principles alone cannot guarantee ethical AI. *Nature Machine Intelligence* **1**, 501-507 (2019).

26 Bietti, E. in *Proceedings of the 2020 conference on fairness, accountability, and transparency.* 210-219.

27 Book, A. *AI Ethics doesn't exist*, <https://towardsdatascience.com/ai-ethics-doesnt-exist-87803ee8fddc> (2020).

28 Tzachor, A., Devare, M., King, B., Avin, S. & Ó hÉigeartaigh, S. Responsible artificial intelligence in agriculture requires systemic understanding of risks and externalities. *Nature Machine Intelligence* **4**, 104-109, doi:10.1038/s42256-022-00440-4 (2022).

29 Soklaski, R., Goodwin, J., Brown, O., Yee, M. & Matterer, J. Tools and Practices for Responsible AI Engineering. *arXiv preprint arXiv:2201.05647* (2022).

30 Meadows, D. H. Leverage points: Places to intervene in a system. (1999).

31 Dorninger, C. *et al.* Leverage points for sustainability transformation: a review on interventions in food and energy systems. *Ecological Economics* **171**, 106570 (2020).

32  Rosengren, L. M., Raymond, C. M., Sell, M. & Vihinen, H. Identifying leverage points for strengthening adaptive capacity to climate change. *Ecosystems and People* **16**, 427-444, doi:10.1080/26395916.2020.1857439 (2020).

33  Ramsey, A. T., Prentice, D., Ballard, E., Chen, L.-S. & Bierut, L. J. Leverage points to improve smoking cessation treatment in a large tertiary care hospital: a systems-based mixed methods study. *BMJ open* **9**, e030066 (2019).

34  Lu, Q., Zhu, L., Xu, X. & Whittle, J. Responsible-AI-by-Design: a Pattern Collection for Designing Responsible AI Systems. *arXiv preprint arXiv:2203.00905* (2022).

35  Arrieta, A. B. *et al.* Explainable Artificial Intelligence (XAI): Concepts, taxonomies, opportunities and challenges toward responsible AI. *Information fusion* **58**, 82-115 (2020).

36  Sen, P. & Ganguly, D. in *Proceedings of the AAAI Conference on Artificial Intelligence.* 2685-2692.

37  Häußermann, J. J. & Lütge, C. Community-in-the-loop: towards pluralistic value creation in AI, or—why AI needs business ethics. *AI and Ethics*, 1-22 (2021).

38  Green, B. Data science as political action: grounding data science in a politics of justice. *Journal of Social Computing* **2**, 249-265 (2021).

39  Powles, J. & Nissenbaum, H. *The Seductive Diversion of 'Solving' Bias in Artificial Intelligence.* , <https://onezero.medium.com/the-seductive-diversion-of-solving-bias-in-artificial-intelligence-890df5e5ef53> (2018).

40  Lauer, D. Facebook's ethical failures are not accidental; they are part of the business model. *AI and Ethics* **1**, 395-403, doi:10.1007/s43681-021-00068-x (2021).

41  Hao, K. *He got Facebook hooked on AI. Now he can't fix its misinformation addiction*, <https://www.technologyreview.com/2021/03/11/1020600/facebook-responsible-ai-misinformation/> (2021).

42  Jobin, A., Ienca, M. & Vayena, E. The global landscape of AI ethics guidelines. *Nature Machine Intelligence* **1**, 389-399 (2019).

43  Kish-Gephart, J. J., Harrison, D. A. & Treviño, L. K. Bad apples, bad cases, and bad barrels: meta-analytic evidence about sources of unethical decisions at work. *Journal of applied psychology* **95**, 1 (2010).

44  Wagner, B. in *Being Profiled: Cogitas Ergo Sum 10 Years of Profiling the European Citizen.* (eds E. Bayamlioğlu, I. Baraliuc, & L. Janssens) 84–88 (Amsterdam University Press, 2018).

45  Nemitz, P. Constitutional democracy and technology in the age of artificial intelligence. *Philosophical Transactions of the Royal Society A: Mathematical, Physical and Engineering Sciences* **376**, 20180089 (2018).

46  Abson, D. J. *et al.* Leverage points for sustainability transformation. *Ambio* **46**, 30-39 (2017).

47  McLennan, S. *et al.* An embedded ethics approach for AI development. *Nature Machine Intelligence* **2**, 488-490, doi:10.1038/s42256-020-0214-1 (2020).

48  Nabavi, E., Daniell, K. A., Williams, E. T. & Bentley, C. M. in *Artificial Intelligence – For Better or Worse* (eds Niels Wouters, Grant Blashki, & Helen Sykes) 157-176 (Future Leaders, 2019).